\documentstyle[prl,aps,psfig,axodraw]{revtex}
\draft

\begin{document}

\newcommand{\met}{\rlap{\,/}E_T}
\newcommand{\be}{\begin{equation}}
\newcommand{\ee}{\end{equation}}
\newcommand{\bear}{\begin{eqnarray}}
\newcommand{\eear}{\end{eqnarray}}
\newcommand{\gsim}{\lower.7ex\hbox{$\;\stackrel{\textstyle>}{\sim}\;$}}
\newcommand{\lsim}{\lower.7ex\hbox{$\;\stackrel{\textstyle<}{\sim}\;$}}

\twocolumn[\hsize\textwidth\columnwidth\hsize\csname
@twocolumnfalse\endcsname

\title{Higgs Boson Decays to CP-odd 
Scalars at the Tevatron and Beyond}

\author{Bogdan A.~Dobrescu$^a$,
Greg Landsberg$^b$ and Konstantin T.~Matchev$^a$}

\address{$^a$ Theoretical Physics Department, 
              Fermi National Accelerator Laboratory, 
              Batavia, IL 60510\\
         $^b$ Department of Physics, Brown University,
              Providence, RI 02912}

\maketitle

\vspace*{.5mm}
\centerline{May 30, 2000}

\begin{abstract}
In extended Higgs models, the Higgs boson may decay into
a pair of light CP-odd scalars, with distinctive
collider signatures. We study the ensuing Higgs signals
at the upgraded Tevatron, considering the subsequent decays of the
scalars into pairs of gluons or photons. For CP-odd scalars
lighter than a few GeV, the Higgs boson manifests itself
as a diphoton resonance and can be discovered up to
masses of a few hundred GeV. For heavier CP-odd scalars
the reach extends at most up to $M_h\sim 120$ GeV.
We also discuss the capabilities of the LHC and lepton colliders
in these channels.

\end{abstract}

\vspace*{.5mm}

\pacs{PACS numbers: 12.60.Rc, 14.80.Cp, 14.80.Mz
\hfill FERMILAB-Pub-99/324-T
}
\vspace*{.5mm}
]

\setcounter{footnote}{0}
\setcounter{page}{1}

The Higgs boson is the only Standard Model (SM) particle
that remains elusive.
The next runs at the Fermilab Tevatron have the potential 
for discovering the Higgs boson for a mass range beyond the
current LEP limit \cite{LEPmh}.
Remarkable efforts \cite{higgsreport}
have been devoted to designing the best
search strategy for a SM-like Higgs boson,
with a large branching into $b$-jets or $W$-bosons.
However, a Higgs boson, $h^0$, with SM-like couplings
to the gauge bosons and fermions, could nevertheless have 
decay modes dramatically different from the SM ones.
The reason is that the SM is likely to be a part of
a more comprehensive theory which may include
an extended Higgs sector. It is then possible
for the Higgs boson to decay into
pairs of other neutral scalars, whenever
they are lighter than half the $h^0$ mass, $M_h$.

An example of such an extended Higgs sector is given by the 
Minimal Composite Higgs Model (MCHM) \cite{MCHM}, which is
based on the top-quark condensation seesaw mechanism \cite{see,saw}.
At low energy, the MCHM includes two composite Higgs doublets
and two gauge-singlet scalars, with the $h^0$ and a CP-odd scalar, 
$A^0$, being the lightest scalar mass eigenstates.
Another interesting example is the Next-to-Minimal
Supersymmetric Standard Model (NMSSM) \cite{HHG}, where the presence 
of two Higgs doublets and a gauge singlet allows a region of
parameter space in which $M_h$ is larger 
than twice the mass $M_A$ of the lightest CP-odd scalar, $A^0$.
Both in the MCHM and NMSSM it is natural to have a light $A^0$
because its mass is controlled by the explicit breaking of
a spontaneously broken $U(1)$ symmetry.
In both models this global $U(1)$ symmetry has a QCD anomaly,
and therefore $A^0$ is an axion.
Note though that various axion searches \cite{PDG} 
place a lower bound on $M_A$, typically in the MeV range,
which requires
explicit $U(1)$ breaking beyond the QCD
anomaly, so that $A^0$ does not solve the strong CP problem.
In other models, such as the chiral supersymmetric 
models \cite{CDM}, or composite Higgs models from extra dimensions 
\cite{Cheng:1999bg}, $h^0$ could also decay into light CP-even scalars.

In this Letter we study the Higgs boson decay
into CP-odd scalar pairs at the upgraded Tevatron. 
We assume the existence of a scalar $A^0$
(we call it ``axion'' for short), 
of mass $M_A < M_h/2$, with a trilinear coupling
\be
\frac{c\,v}{2} h^0 A^0 A^0 ~,
\label{coupling}
\ee
where $v \approx 246$ GeV is the electroweak symmetry breaking
scale, and $c$ is a model-dependent dimensionless parameter,
which can be as large as ${\cal O}(1)$.
 
\begin{figure}[h!]
\centerline{\psfig{figure=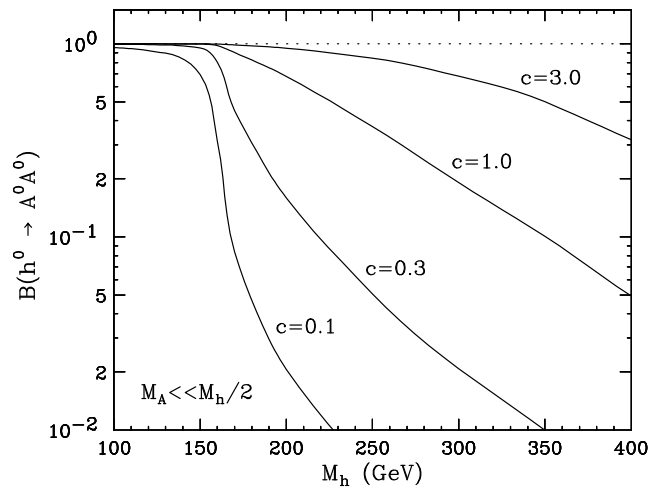,height=2.5in}}
\begin{center}
\parbox{5.5in}{
\caption[] {\small Branching ratio of the Higgs boson
into axion pairs, as a function of $M_h$,
for $M_A\ll M_h/2$ and several values of $c$.
\label{br_higgs}}}
\end{center}
\end{figure}

The Higgs width into a pair of axions is
\begin{equation}
\Gamma(h^0\rightarrow A^0A^0)\ =\ 
{c^2\, v^2\over 32\pi M_h}\ 
\left(1-4{M_A^2\over M_h^2}  \right)^{1/2}.
\label{width}
\end{equation}
The decay to axion pairs can be
essential for Higgs boson searches in collider experiments.
This is illustrated in Fig.~\ref{br_higgs},
where we plot the Higgs boson branching ratio to axions,
$B(h^0\rightarrow A^0A^0)$,
versus $M_h$, for $M_A\ll M_h/2$ and several values of $c$.
For $M_h$ below the $WW$ threshold,
the dominant SM decay of the Higgs boson
$h^0\rightarrow b\bar{b}$ has a very small width,
and is therefore susceptible to the presence of new
physics beyond the SM, {\it e.g.} the interaction (\ref{coupling}).
The decay to axions would then dominate over
$h\rightarrow b\bar{b}$ for values of $c$ as small as
$\sim 0.02  
(M_h/100\ {\rm GeV})$.
We also see that even above the $WW$ threshold,
$h^0\rightarrow A^0A^0$ competes with
$h^0\rightarrow WW$, provided $c\sim {\cal O}(1)$.

{\it Experimental signatures of the Higgs boson}.---
The final state signatures for $h^0\rightarrow A^0A^0$ searches
will depend on the subsequent decays of the axions,
which are quite model-dependent.
Here we will concentrate on the case where the 
axion couplings to light fermions are negligible. 
Such a situation may arise in the MCHM, {\it e.g.\/}
when only one Higgs doublet is mainly responsible for
electroweak symmetry breaking and fermion masses
(type I two-Higgs-doublet model with large $\tan\beta$ \cite{HHG}).
In the MCHM, however, the axion has a large coupling to a heavy 
vector-like quark, $\chi$, whose charges under the SM gauge group 
are the same as for the right-handed top quark $t_R$.
This coupling allows one-loop decays of the 
axion into gluon or photon pairs. The case where the axion couplings
to light fermions are significant ({\it e.g.}, in the NMSSM)
provides interesting experimental signatures as well,
but we leave its investigation for future studies. 

The constraints on 
a light CP-odd scalar with small couplings to SM fermions 
are loose \cite{MCHM}.
For example, no constraints on the axion mass have been set from
$Z\rightarrow A^0\gamma$ at LEP \cite{ZAgamma},
from direct $A^0$ production through
gluon fusion via a $\chi$ loop at the Tevatron \cite{DRRC},
from fits to the electroweak data,
or from meson decays \cite{HHG}.
The relevant lower bounds on $M_A$ come from beam dump 
experiments \cite{FT}, in the MeV range, and from star 
cooling rates, $M_A \gsim 0.2$ MeV \cite{PDG}.

In the MCHM, the two Higgs doublets and
two gauge-singlet scalars arise as bound states
of the top-bottom left-handed doublet
or $\chi_L$, with $\chi_R$ or $t_R$.
At scales below a few TeV this is an explicit,
renormalizable theory with scalars that {\it appear}
to be fundamental. The Higgs potential has been
analyzed in detail in \cite{MCHM,saw}.
For the purpose of studying Higgs boson decays in the MCHM
it is however sufficient to introduce the following simpler model. 
Consider the SM with the addition of a gauge singlet complex scalar
$S$ and the vector-like quark, $\chi$.
Besides the kinetic terms and Higgs Yukawa couplings,
the following terms are present in the Lagrangian:
\begin{equation}
{\cal L} = \xi S \overline{\chi}_L \chi_R +
{\rm h.c.} - V(H,S)~,
\end{equation}
where $\xi$ is a Yukawa coupling, and the scalar potential is
\begin{eqnarray}
V(H,S) &=& {\lambda_H\over2} \left(H^\dagger H\right)^2 
+ {\lambda_S\over2} \left(S^\dagger S\right)^2 
+ \lambda_0 H^\dagger H S^\dagger S \nonumber \\
&+& M_H^2 H^\dagger H + M_S^2 S^\dagger S
+ C_S \left(S + S^\dagger\right) ~.
\label{potential}
\end{eqnarray}
We assume $M_H^2 < 0$, and $\lambda_H\lambda_S > \lambda_0^2$.
In the limit where the coefficient $C_S$ of the tadpole term vanishes, 
the effective potential has a global $U(1)$ symmetry spontaneously
broken by the vacuum expectation value $\langle S \rangle$ of $S$.
The associated axion, $A^0$, is part of the 
singlet $S$, and due to the tadpole term has a mass given by
$\sqrt{|C_S/\langle S \rangle|}$. 
Note that in the MCHM the tadpole term is generated by a tree level
$\chi$ mass, and there is mixing between $\chi$ and $t$ once the
electroweak symmetry is broken. These details are not relevant for the 
present study, but the coupling of the axion to $\chi$ is essential 
since otherwise the axion would be stable and the 
Higgs boson would decay invisibly \cite{Datta}.

The $h^0 A^0A^0$ coupling may be easily computed in the SM+singlet
model, with the result
\be
c =  -\ \sqrt{2} \lambda_S \frac{\langle S\rangle}{v} \sin\theta 
- \lambda_0 \cos \theta ~,
\ee
where $\theta$ is the mixing angle between the two CP-even neutral
scalars,
\be
\tan 2\theta \simeq \frac{-\,2\sqrt{2}\lambda_0 v 
\langle S\rangle}{2\lambda_S\langle S\rangle^2 - \lambda_H v^2 
- C_S/\langle S\rangle}~.
\ee
For a range of values of the six parameters from Eq.~(\ref{potential}),
the Higgs decay to axions is important.

The effective axion couplings to gluons and photons, induced at 
one-loop by the $\chi$ quark, are given by
\begin{equation}
\frac{-\sqrt{2}}{16\pi\langle S\rangle }
A^0 \epsilon^{\mu\nu\rho\sigma}\left(\alpha_s 
{\bf G}_{\mu\nu}{\bf G}_{\rho\sigma} + 
N_c e_\chi^2 \alpha F_{\mu\nu}F_{\rho\sigma}
\right) ~,
\label{effective}
\end{equation}
where $\alpha_s$ ($\alpha$) is the strong coupling (fine structure) constant,
${\bf G}_{\mu\nu}$ ($F_{\mu\nu}$) is the gluon (photon) field
strength,
$N_c=3$ is the number of colors and $e_\chi=2/3$ is the electric
charge of the vector-like quark.
Naively, the dominant axion decay mode is to a pair of gluons. 
However, the gluons must then hadronize, and for
light axions the number of open channels is very limited.
Indeed, the two-body decays $A^0 \rightarrow \pi\pi, \pi^0\gamma$
are forbidden by CP invariance and angular momentum conservation,
while the three body decays
$A^0 \rightarrow \pi^0\gamma\gamma, 2\pi^0\gamma, \pi^+\pi^-\gamma$
are significantly suppressed by phase space
(note that C- and P-parity need not be separately conserved
in axion decays). Therefore, for $M_A \lsim 3m_\pi\approx 405$ MeV
the branching ratio to two photons is close to 1. 
The axion decay width due to the two photon effective coupling 
(\ref{effective}) is given by
\begin{equation}
\Gamma(A^0 \rightarrow \gamma\gamma)
\simeq \frac{\alpha^2 M_A^3 }{72\pi^3 \langle S\rangle^2}~.
\label{Gamma}
\end{equation}
For $M_A \gsim 0.5$ GeV the isospin-violating decay modes
$A^0 \rightarrow 3\pi$ open up and
start to compete with $A^0 \rightarrow \gamma\gamma$.
Due to QCD uncertainties,
it is quite difficult to estimate their exact width,
but recall that the measured branching fractions of
$\eta$ decays into $\gamma\gamma$,
$3\pi^0$, and $\pi^+\pi^-\pi^0$ are given 
by $39\%$, $32\%$, and $23\%$, respectively \cite{PDG}.
Since $\eta$ and $A^0$ have the same quantum numbers,
the $A^0 \rightarrow \gamma\gamma$ decay mode is likely to
be significant even for $M_A$ of order 1 GeV.
Fortunately, the study of the $h^0\rightarrow A^0A^0$ decay
in this $M_A$ range does not require a
precise determination of the axion branching ratios,
as we explain below.

\begin{figure}[ht]
\unitlength=0.7 pt
\SetScale{0.7}
\SetWidth{0.7}      
\normalsize    
{} \qquad\allowbreak
\begin{picture}(99,99)(0,0)
\Text(175,53)[b]{$A^0$}
\Text(125,53)[b]{$A^0$}
\DashLine(150.0,50.0)(200.0,50.0){3.0} 
\DashLine(150.0,50.0)(100.0,50.0){3.0} 
\Vertex(150,50){3}
\Text(150,56)[b]{$h^0$}
\Line(200.0,50.0)(250.0,53.0)
\Line(250.0,47.0)(200.0,50.0) 
\Text(220,58)[b]{$\gamma$}
\Text(220,43)[t]{$\gamma$}
\Line(100.0,50.0)(50.0,53.0)
\Line(50.0,47.0)(100.0,50.0) 
\Text(80,58)[b]{$\gamma$}
\Text(80,43)[t]{$\gamma$}
\Text(150,30)[t]{$(a)$}
\end{picture} \ 

\vspace{-0.6cm}
\unitlength=0.7 pt
\SetScale{0.7}
\SetWidth{0.7}      
\normalsize    
{} \qquad\allowbreak
\begin{picture}(99,99)(0,0)
\Text(175,53)[b]{$A^0$}
\Text(125,53)[b]{$A^0$}
\DashLine(150.0,50.0)(200.0,50.0){3.0} 
\DashLine(150.0,50.0)(100.0,50.0){3.0} 
\Vertex(150,50){3}
\Text(150,56)[b]{$h^0$}
\Line(200.0,50.0)(250.0,54.0)
\Line(200.0,50.0)(250.0,50.0) 
\Line(200.0,50.0)(250.0,46.0) 
\Text(220,58)[b]{3$\pi^0$}
\Line(100.0,50.0)(50.0,54.0)
\Line(100.0,50.0)(50.0,50.0)
\Line(100.0,50.0)(50.0,46.0)
\Text(80,58)[b]{3$\pi^0$}
\Text(280,65)[b]{6$\gamma$}
\Line(250.0,54.0)(300.0,58.0)
\Line(250.0,54.0)(300.0,54.0)
\Line(250.0,50.0)(300.0,51.5) 
\Line(250.0,50.0)(300.0,48.5) 
\Line(250.0,46.0)(300.0,46.0) 
\Line(250.0,46.0)(300.0,42.0) 
\Text(20,65)[b]{6$\gamma$}
\Line(0.0,58.0)(50.0,54.0)
\Line(0.0,54.0)(50.0,54.0)
\Line(0.0,51.5)(50.0,50.0)
\Line(0.0,48.5)(50.0,50.0)
\Line(0.0,46.0)(50.0,46.0)
\Line(0.0,42.0)(50.0,46.0)
\Text(150,30)[t]{$(b)$}
\end{picture} \ 
\begin{center}
\vspace*{-0.2cm}
\parbox{5.5in}{
\caption[] {\small Higgs boson decay topology
into a ``diphoton'' final state
with (a) prompt and (b) cascade photons.
\label{decay}}}
\end{center}
\end{figure}
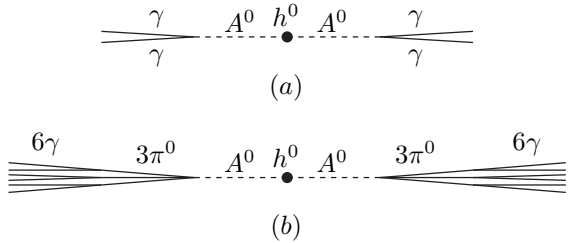

The Higgs decays are quite peculiar in the scenario discussed here.
Since the LEP \cite{LEPdiphotons} and Tevatron \cite{FNALdiphotons}
limits would generally apply, we only consider 
the range of Higgs masses above $\sim 100$ GeV.
Because of the relatively heavy parent mass, each axion will be
produced with a significant boost, and will decay into a pair
of almost colinear photons. For $M_A \lsim 0.025M_h$,
they will not be resolved in the electromagnetic calorimeter
and will be reconstructed as a single photon. As a result, the 
$h \rightarrow A^0A^0 \rightarrow 4\gamma$ mode will appear 
in the detector as a diphoton signature, as shown in Fig.~\ref{decay}a.
\vspace{-0.7cm}
\vspace*{-0.5cm}

The interesting twist is that the $A^0 \rightarrow 3\pi^0$
decay mode will have the same signature in the detector, because
$\pi^0$'s decay promptly into photons. This situation is
illustrated in Fig.~\ref{decay}b.
We can continue this line of argument for even higher $M_A$.
If $M_A \gsim 1$ GeV, the $\omega\gamma$, $\rho\gamma$,
and $\eta\pi\pi$ axion decay modes become relevant.
Using the measured branching fractions of $\eta^\prime$ 
into these states, and the subsequent $\eta, \rho, \omega$ 
decays, we find that the $\eta^\prime$ branching fraction for final 
states with only photons is roughly 17\%.
Given the similarities between $\eta^\prime$ and the axion,
we expect the probability that $A^0$ is reconstructed as a photon is
of order 20\% when $M_A \sim 1$ GeV.

For $M_A \gsim 2$ GeV, new isospin-conserving
modes with large branching fractions open up: 
$A^0 \rightarrow \rho\rho$, $\omega\omega$, $a_0\pi^0$, $KK\pi^0$, etc.
(isospin-violating decays, such as $\eta\eta\pi$,  
are suppressed.)
Even then, some of these mesons have large branching fractions
into states which subsequently decay 
into photons, {\it e.g.}, $a_0 \rightarrow \eta\pi^0$ \cite{PDG},
yielding the same diphoton signature for $h^0$.
Only when $M_A$ is increased above several GeV do
the decay products of $h^0$ look more
like QCD jets instead of photons.

An important issue is whether the $A^0$ decays promptly. 
Using Eq.~(\ref{Gamma}), we can estimate its decay length
\be 
L_A \approx
4 \; {\rm mm} \; {M_h\langle S \rangle^2 \over 
\left(100\ {\rm GeV}\right)^3}
\left( {1\ {\rm GeV}\over M_A} \right)^4
\ .
\ee
For $M_A \lsim 100$ MeV the axion decays occur most of the time
outside the detector, so the Higgs boson decay is invisible
\cite{invisible}. In the following we shall instead concentrate
on the mass range $M_A \gsim 200$ MeV, 
where the axion decays before reaching the electromagnetic calorimeter.

\begin{figure}[t!]
\centerline{\psfig{figure=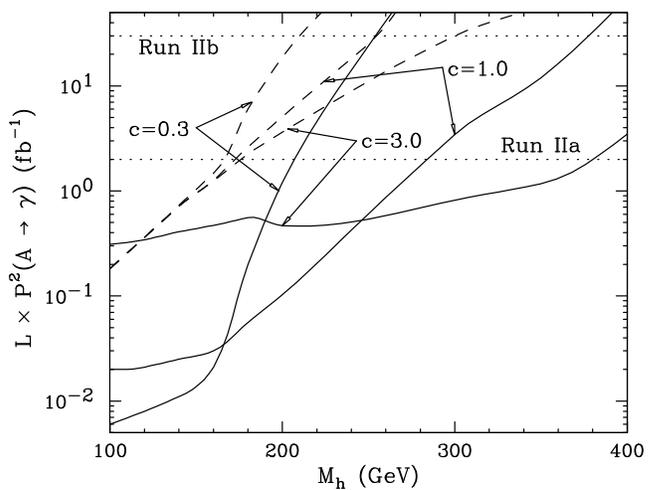,height=2.5in}}
\begin{center}
\parbox{5.5in}{
\caption[] {\small The Tevatron reach at 95\% C.L. in the inclusive
diphoton channel (solid) and $\ell2\gamma\met$ channel
(dashed), as a function of $M_h$, for three different
values of $c$.
\label{reach_gg}}}
\end{center}
\end{figure}

{\it Tevatron reach for light axions}.---
For $M_A$ less than a few GeV, we have studied the Tevatron reach 
in the diphoton channel, following the analysis of Ref. \cite{diphotons}.
We used {\textsc PYTHIA} \cite{PYTHIA} for event generation
and {\textsc SHW} \cite{SHW} with minor modifications
\cite{modifications} for detector simulation.
The Higgs boson is produced predominantly via gluon fusion,
and the inclusive diphoton channel (with an optimized
cut on the invariant diphoton mass, $m_{\gamma\gamma}$)
provides the best sensitivity \cite{diphotons}.
The reach is shown in Fig.~\ref{reach_gg},
where we plot the product $L \times P^2(A\rightarrow\gamma)$
as a function of $M_h$. Here $L$ is
the total integrated luminosity required
for a 95\% confidence level (C.L.) exclusion, and $P(A\rightarrow\gamma)$
is the ($M_A$-dependent) probability that an axion is reconstructed as
a photon, including the branching fractions of cascade
decays of axions into multi-photon final states. 
We see that already run IIa with $2\ {\rm fb}^{-1}$
will be probing a range of Higgs boson masses well beyond the reach
via conventional SM searches \cite{higgsreport}.
Note the reversed ordering of the lines of
constant $c$ at low and high $M_h$.
Below the $WW$-threshold, the Higgs boson width, $\Gamma_h$,
is dominated by the decay (\ref{width}) into axions,
hence a larger $c$ requires
a softer $m_{\gamma\gamma}$ cut and leads to higher background.
Above the $WW$-threshold, where $\Gamma_h$ is dominated
by the SM decays, a larger $c$
is beneficial due to the increased
$B(h\rightarrow A^0A^0)$ (see Fig.~\ref{br_higgs}).
\vspace*{-0.5cm}

For light axions, associated $Wh/Zh$ production can give
alternative, very clean signatures: $\ell\gamma\gamma\met$
and $\ell^+\ell^-\gamma\gamma$.
Considering only leptonic decays of the $W/Z$, we expect
$Wh$ to give a better reach, because of the larger
leptonic branching fraction. Requiring events with
$\met>20$ GeV, at least one lepton with $p_T(\ell)>20$ GeV,
and at least two photons with $p_T(\gamma)>20$ GeV
and $|m_{\gamma\gamma}-M_h|<\Gamma_h$,
we find the following parametrization of the
signal efficiency: $\varepsilon=0.32-0.07(M_h/100\ {\rm GeV})$.
The main background is $W\gamma\gamma$ and we estimated
it using COMPHEP \cite{comphep} to be less
than 1 event after cuts in 30 ${\rm fb}^{-1}$.
The corresponding reach 
is shown in Fig.~\ref{reach_gg} (dashed lines).
We see that although this channel gives a smaller absolute
reach than the inclusive $2\gamma$ channel,
it will still enable the Tevatron to probe 
Higgs masses below $WW$ threshold, for a wide
range of values of $c$.

{\it Tevatron reach for heavy axions}.---
If $M_A$ is above a few GeV, the decay products from
$A\rightarrow gg$ look like QCD jets,
and the background is large. To make matters worse, at the
parton level $A^0\rightarrow gg$ dominates over
$A^0\rightarrow \gamma\gamma$ by a factor
of $(N_c^2-1)\alpha_s^2/(4N_c^2\alpha^2e_\chi^4) \approx 265$.
Typically, for $M_A\gsim 20$ GeV,
the resulting two gluon jets are separated well enough
and can both be individually reconstructed.
Unlike the SM, the final states from Higgs production
contain no $b$-jets, but instead mostly gluon jets
and very rarely photons. 
Since at present it is practically impossible
to distinguish between gluon and quark jets,
these searches appear quite challenging.

Gluon fusion ($gg\rightarrow h\rightarrow A^0A^0$)
can lead to $4j$, $2j2\gamma$ or $4\gamma$
events. The $4j$ channel suffers from insurmountable
backgrounds, while the $4\gamma$ channel has a
tiny branching ratio, and will escape detection even
in run IIb with 30 ${\rm fb}^{-1}$.
The $2j2\gamma$ channel is similar to the searches for a ``bosonic
Higgs'' at the Tevatron \cite{FNALdiphotons,diphotons,MW}.
Using the same cuts as for the inclusive
$\gamma\gamma+N{\rm jets}$ channels of
Ref.~\cite{diphotons}, we found no reach in run IIb.

Searches for associated Higgs boson production
also appear very difficult. 
The 6-jet final states from all-hadronic decays of
the $W$ ($Z$) and $A^0$ have large QCD backgrounds.
Leptonic decays of the $W$ and $Z$, combined with
$h\rightarrow 4j$ give $\ell4j\met$
and $\ell^+\ell^-4j$, respectively.
The backgrounds are large and again no sensitivity in
run II is expected. Finally, requiring that
one axion decays to $\gamma\gamma$, and
leptonic decays of the $W$ or $Z$, we get the
relatively clean $\ell 2\gamma \met + X$ and
$\ell^+\ell^-2\gamma + X$ final states.
Because of the $B(A^0\rightarrow\gamma\gamma)$ suppression,
the signal rates are too small to be observed in run IIa,
but run IIb might be able to explore the mass range
$100\lsim M_h\lsim 120$ GeV.

{\it Discovery prospects at the LHC and future lepton colliders}.---
It is interesting to contemplate the capabilities of
future colliders for our scenario.  The LHC has enormous potential
for such Higgs boson searches.
Just like at the Tevatron, one will have to concentrate on
the cleanest channels. But the much larger signal rate
will allow one to look for the photonic decays of heavy axions,
which were severely limited by statistics at the Tevatron.
Preliminary studies show that even the $2j 2\gamma$ channel,
swamped by the QCD background at the Tevatron, is sensitive to
a large range of Higgs boson and axion masses at the LHC, due to
the improved energy resolution of the LHC detectors and
the enhanced signal cross section.

A high energy lepton collider (such as the NLC) would be
ideally suited for unravelling a non-standard Higgs sector,
like the one discussed in this Letter,
particularly if $M_A$ is bigger than a few GeV and
$P(A \to \gamma)$ is very small.
Notice that the jet-rich channels are the best to look
for at a lepton collider, since they have the largest branching
ratios. In this sense, lepton colliders are complementary to
hadron machines, where these channels suffer
from large backgrounds. We also expect that LEP-II will be
able to probe Higgs masses up to its kinematic reach,
once a dedicated search is done. We therefore urge the
LEP collaborations to present Higgs mass limits 
with the data selection optimized for the discussed signatures.

In conclusion, we have considered several novel Higgs boson
discovery signatures, arising from the decay $h^0\rightarrow A^0A^0$,
present in many extended Higgs sector models.
Quite ironically, the best reach at the Tevatron is obtained
for light axions, where the Higgs boson often
decays to two jets, each of which ``fakes'' a photon.

{\it Acknowledgements.}---
We thank A.~Falk, J.~Bagger, H.~Haber, G.~Hiller, H.~Logan, J.~Lykken,
S.~Mrenna, U.~Nierste, and D.~Rainwater for comments.
This research was supported in part by the
U.S.~Department of Energy (DOE) Grant DE-FG02-91ER40688.
Fermilab is operated under DOE contract DE-AC02-76CH03000.

\end{document}